\documentclass[twocolumn,showpacs,preprintnumbers,amsmath,amssymb]{revtex4}
\usepackage{graphicx}
\usepackage{amsmath}
\usepackage{bm}
\usepackage{color}

\newcommand{\bald}[1]{{\bf #1}}

\usepackage{graphicx}
\usepackage{dcolumn}
\usepackage{bm}

\begin{document}


\title{Mach cones in the quark-gluon plasma: Viscosity, speed of sound, and effects of finite source structure}

\author{R. B. Neufeld}
   \affiliation{Department of Physics, Duke University, Durham, North Carolina 27708, USA}
    \email{rbn2@phy.duke.edu}

\date{\today}

\begin{abstract}
I use the space-time distribution of energy and momentum deposited by a fast parton traversing a perturbative quark-gluon plasma as a source term for the linearized hydrodynamical equations of the medium.  A method of solution for the medium response is presented in detail.  Numerical results are given for different values of the shear viscosity to entropy density ratio, $\eta/s$, and speed of sound, $c_s$.  Furthermore, I investigate the relevance of finite source structure by expanding the source term up to first order in gradients of a $\delta$ function centered at the fast parton and comparing the resulting dynamics to that obtained with the full source.  It is found that, for the source term used here, the medium response is sensitive to the finite source structure up to distances of several fm from the source parton.
\end{abstract}


\maketitle

\begin{figure}
\centerline{
\includegraphics[width = 0.95\linewidth]{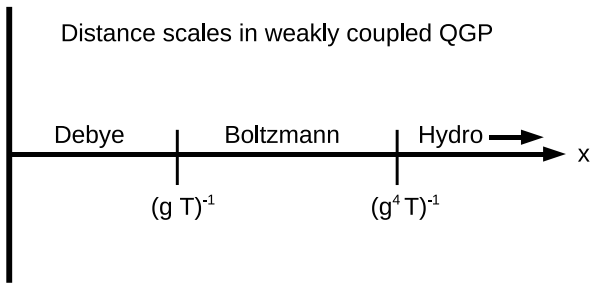}
}
\caption{Some of the distance scales relevant to the hydrodynamic response of a weakly coupled QGP to fast partons.}
\label{qcdscales}
\end{figure}

\section{Introduction}
A relatively new and exciting problem in {\it quark-gluon plasma} (QGP) physics is to determine the response of the medium to the passage of a fast parton.  Fast partons are created by hard transverse scattering in the early moments of a heavy-ion collision and have long been considered a useful probe in understanding the QGP.  The primary emphasis has focused on the process of {\it jet quenching} in which fast partons lose energy and momentum by interacting with the surrounding medium (see, for instance, Refs. \cite{Wang:1991xy,Baier:1996sk,Zakharov:1997uu,Gyulassy:2000fs,Guo:2000nz,Baier:2000mf,Armesto:2004ud,Jacobs:2004qv}).  Recently, the question of how the energy and momentum deposited by the fast parton affects the bulk behavior of an evolving QGP has gained attention (see, e.g., Refs. \cite{CasalderreySolana:2004qm,Stoecker:2004qu,Satarov:2005mv,Ruppert:2005uz,Renk:2005si,Chaudhuri:2005vc,Friess:2006fk,Chesler:2007sv,Betz:2008js,Neufeld:2008fi}).  Interest in understanding the medium's response to the passage of a fast parton has been spurred on by experimental measurements at the Relativistic Heavy-Ion Collider (RHIC) \cite{Adams:2005ph,Adler:2005ee,Ulery:2007zb} of hadron correlation functions which suggest the fast parton may produce a propagating Mach cone in the medium.

There is strong evidence \cite{Ludlam:2005gx,Gyulassy:2004zy} that the matter produced at RHIC obeys the hydrodynamic assumption of local thermal equilibrium.  For this reason, the common theoretical approach to examining the QGP's response to a fast parton has been to treat it as a source of energy and momentum coupled to the hydrodynamic equations of the medium.  This makes sense provided the medium maintains local thermal equilibrium following the passage of a fast parton.  Assuming the medium does respond hydrodynamically to a fast parton then raises the question of what the distribution of energy and momentum deposited is.  It has been observed \cite{CasalderreySolana:2004qm,Chaudhuri:2005vc,Betz:2008js} that the medium's response to fast partons is sensitive to the specific form of energy and momentum deposition, creating the need for a hydrodynamic source term derived from first principles.

It is instructive to consider the mechanism of energy and momentum deposition and the different scales involved.  In a quantum chromodynamic (QCD) plasma fast partons interact with the medium at a distance scale of the order of the inverse Debye mass, $(m_D)^{-1}$.  This interaction creates a disturbance, which in turn interacts with the surrounding medium, creating a new disturbance at some larger distance scale. The new disturbance again interacts with the surrounding medium, and eventually the initial disturbance propagates outward to some arbitrarily large distance scale.  At distances much greater than the mean free path, $\Lambda_f$, the medium's response to the initial disturbance can be accurately described by hydrodynamics.  Thus an effective QCD hydrodynamic source term should include the medium's response up to distances of order $\Lambda_f$, at which point the system evolves hydrodynamically.  Whether the plasma is weakly or strongly coupled, the initial energy and momentum deposition occurs at a distance scale of the order of the inverse Debye mass, although the specific value of the Debye mass depends on the strength of the coupling.  However, in a strongly coupled plasma, the concept of a mean free path loses meaning, and instead the de Broglie wavelength sets the minimum scale at which the hydrodynamical description is valid.  In general, the application of hydrodynamics is valid on shorter distance scales for more strongly coupled mediums.

Recently, Neufeld \cite{Neufeld:2008hs} presented a derivation of the hydrodynamic source term expected from a fast parton moving through a perturbative QGP, both with and without including the effect of color screening, by including the medium response at a distance scale of the order $(m_D)^{-1}$.  Using the unscreened, relativistic form of this source term coupled to the linearized hydrodynamical equations of the medium the authors of Ref. \cite{Neufeld:2008fi} showed that the medium response includes a propagating sound wave with the shape of a Mach cone and a diffusive wake.  In this work, I will use a slightly modified form of the relativistic limit of this source term in the linearized hydrodynamical equations of the medium.  A detailed solution of the equations of motion will be presented along with the resulting dynamics for a range of values of the shear viscosity to entropy density ratio, $\eta/s$, and speed of sound, $c_s$.  I will also expand the source term up to first order in a series of gradients of a $\delta$ function and compare to the full result, in an effort to understand the relevance of finite source structure.  It will be shown that the medium response is sensitive to the finite source structure up to distances of several fm from the source parton for the source term used here.

In a weakly coupled QCD plasma at high temperature $T$, the inverse Debye mass is of order $(g T)^{-1}$, where $g$ is the running coupling, whereas the transport mean free path is of order $(g^4 T)^{-1}$ \cite{Arnold:2002zm}.  The medium's response to disturbances at distance scales between $(g T)^{-1}$ and $(g^4 T)^{-1}$ is accurately described by the Boltzmann equation (see Fig. \ref{qcdscales}).  The hydrodynamic source term examined here includes the medium's response at a distance scale of the order of the inverse Debye mass, at which point the medium's hydrodynamic response is invoked.  This is a simplification of the true QCD evolution, where the medium's evolution between $(g T)^{-1}$ and $(g^4 T)^{-1}$ should be described by the Boltzmann equation, after which the hydrodynamic response can be invoked.  However, in the QGP produced at RHIC, it is likely that the mean free path is comparable in size to the inverse Debye mass (compare, for instance, Refs. \cite{Kaczmarek:2005ui} and \cite{Molnar:2001ux}).  Thus, from a phenomenological point of view, physics at a distance scale of the order $(m_D)^{-1}$ may be relevant to the specific structure of the medium's hydrodynamic response to fast partons.  However, this also suggests the QGP produced at RHIC may be strongly coupled, limiting the application of perturbation theory.

The paper is structured as follows.  In Sec.\ref{hydrosect} I consider the general form of linearized hydrodynamics with a source term.  I then introduce the specific source term studied here.  In Sec.\ref{deltamom} the source term is expanded in terms of gradients of a $\delta$ function up to first order.  Both the full source term and the $\delta$ function expanded form are then Fourier transformed into momentum space.  In Sec.\ref{solving} the hydrodynamic equations are solved in terms of a one-dimensional numerical integration.  In Sec.\ref{randd} results are given for both the full source and the $\delta$ function expanded version, as well as different values of the viscosity and speed of sound.  In what follows I choose units such that $\hbar = c = k_b = 1$.

\section{Linearized Hydrodynamics With a Source Term}\label{hydrosect}

The first-order hydrodynamical equations for a medium with nonzero shear viscosity $\eta$ in the presence of a source term $J^\nu$ are given by
\begin{equation}\label{sourcehydro}
\partial_\mu T^{\mu \nu} = J^{\nu},
\end{equation}
where $T^{\mu \nu}$ is the energy-momentum tensor of the system.  If one assumes that the energy and momentum density deposited by the fast parton is small compared to the equilibrium energy density of the medium, the hydrodynamical equations [Eq. (\ref{sourcehydro})] can be linearized.  Defining $T^{\mu\nu} = T_0^{\mu\nu} + \delta T^{\mu\nu}$, where $\delta T^{\mu\nu}$ is the perturbation of the energy-momentum tensor resulting from the source in an otherwise static medium, one has
\begin{equation}\label{lin_source}
\partial_\mu \delta T^{\mu\nu} = J^\nu ,
\end{equation}
where $\partial_\mu T_0^{\mu \nu} = 0$ and $\delta T^{\mu\nu}$ is given by \cite{CasalderreySolana:2004qm}
\begin{equation}\label{mom_space_tensor}
\begin{split}
&\delta T^{00} \equiv \delta \epsilon \text{,    }\delta T^{0i} \equiv \bald{g}, \\
& \delta T^{ij} = \delta_{ij} c_s^2 \delta \epsilon - \frac{3}{4} \Gamma_s\left(\partial^i g^j + \partial^j g^i - \frac{2}{3}\delta_{ij} \bald{\nabla}\cdot\bald{g}\right).
\end{split}
\end{equation}
In Eqs. (\ref{mom_space_tensor}) $c_s$ denotes the speed of sound, $\Gamma_s \equiv \frac{4 \eta }{3(\epsilon_0 + p_0)} = \frac{4 \eta }{3 s T}$ is the sound attenuation length, and $\epsilon_0$ and $p_0$ are the unperturbed energy density and pressure, respectively.

By introducing the general rule for Fourier transforms
\begin{equation}\label{generalrule}
F({\bf x},t) = \frac{1}{(2 \pi)^4}\int d^3 k \int d \omega \,e^{i \bald{k}\cdot\bald{x} - i \omega t} F({\bf k},\omega),
\end{equation}
the equations given by Eq. (\ref{lin_source}) are written in momentum space as
\begin{eqnarray}\label{hydro_1}
J^0 &=& -i\omega \delta \epsilon + i \bald{k}\cdot\bald{g}  \\
\label{hydro_2}
\bald{J} &=& -i \omega \bald{g} + i \bald{k} c_s^2 \delta \epsilon + \frac{3}{4} \Gamma_s \left(k^2\bald{g} + \frac{\bald{k}}{3}( \bald{k}\cdot\bald{g})\right).
\end{eqnarray}
Solving for $\bald{k}\cdot\bald{g}$ in Eq. (\ref{hydro_2}) allows for $\delta \epsilon$ to be determined from Eq. (\ref{hydro_1}):
\begin{equation}
\bald{k}\cdot\bald{g} = k g_L = \frac{\bald{k}\cdot\bald{J} - i k^2 c_s^2 \delta \epsilon}{-i\omega + \Gamma_s k^2}
\end{equation}
and hence
\begin{equation}\label{eps}
\delta\epsilon ({\mathbf k},\omega) = \frac{i k J_L({\mathbf k},\omega)  + J^0({\mathbf k},\omega)(i \omega -  \Gamma_s k^2)}{\omega^2 -  c_s^2 k^2 + i \Gamma_s \omega k^2},
\end{equation}
where the source and perturbed momentum density vectors are divided into transverse and longitudinal parts: ${\mathbf g} = \hat{\mathbf k} g_L + {\mathbf g}_T$ and ${\mathbf J} = \hat{\mathbf k} J_L + {\mathbf J}_T$, with $\hat{\mathbf k}$ denoting the unit vector in the direction of ${\mathbf k}$.  Similarly, one has from Eqs. (\ref{hydro_1}) and (\ref{eps})
\begin{equation}
k g_L = -i J^0 + \omega \delta \epsilon
\end{equation}
yielding
\begin{equation}\label{gl}
{\mathbf g}_L ({\mathbf k},\omega) = \hat{\mathbf k} g_L = \frac{ i \omega \hat{\mathbf k} J_L({\mathbf k},\omega)+ i c_s^2 {\mathbf k} J^0({\mathbf k},\omega)}{\omega^2 -  c_s^2 k^2 + i \Gamma_s \omega k^2}.
\end{equation}
The transverse part of $\bald{g}$ can be obtained from Eq. (\ref{hydro_2}).  The calculation is simplified by considering that any part of $\bald{g}$ proportional to $\bald{k}$ is a part of $\bald{g}_L$.  This leaves
\begin{equation}\label{gt}
\bald{g}_T({\mathbf k},\omega) = \bald{g} - \bald{g}_L = \frac{i{\mathbf J}_T({\mathbf k},\omega)}{\omega +  \frac{3}{4}i \Gamma_s k^2}.
\end{equation}

Equation (\ref{gt}) is a diffusion equation and the quantity ${\mathbf g}_T$ is interpreted as diffusive momentum density generated by the fast parton.  Equations (\ref{eps}) and (\ref{gl}) describe damped sound waves propagating at speed $c_s$: it follows that $\delta\epsilon$ and ${\mathbf g}_L$ are interpreted as the energy and momentum density carried by sound generated by the fast parton.  The importance of the explicit form of the source term can be readily seen.  In a homogeneous medium symmetries ensure that the source vector, $\bald{J}$, can be written generally as
\begin{equation}\label{generalj}
\bald{J}(\bald{x},t) = \bald{u}\, p(r) + \bald{\nabla}q(r)
\end{equation}
where $\bald{u}$ is the velocity of the source particle and $p(r)$ and $q(r)$ are scalar functions of the (possibly Lorentz-contracted) magnitude $r = \sqrt{(\bald{x} - \bald{u} t)^2}$.  If we instead write Eq. (\ref{generalj}) in momentum space we have
\begin{equation}\label{fourgen}
\bald{J}(\bald{k},\omega) = \bald{u} \int d^4 x\,e^{i k \cdot x} p(r) + i \bald{k}\int d^4 x\,e^{i k \cdot x} q(r)
\end{equation}
and ${\mathbf J}_T$ is found to be
\begin{equation}
\begin{split}\label{jt}
{\mathbf J}_T (\bald{k},\omega) &= \bald{J} - \frac{\bald{k}(\bald{k}\cdot\bald{J})}{k^2} \\
&= \left(\frac{\bald{u}k^2 - \bald{k}(\bald{k}\cdot\bald{u})}{k^2}\right)\int d^4 x\,e^{i k \cdot x} p(r)
\end{split}
\end{equation}
so that $q(r)$ does not contribute to ${\mathbf J}_T$.  If one chooses the source [Eq. (\ref{generalj})] such that $p(r) = 0$ then there is no excitation of the diffusive momentum density.  It is clear that the hydrodynamics of the system are sensitive to the specific form of the source term.  In particular, terms that are in the form of a gradient only generate sound.

As mentioned previously, in this work I use the source term derived in Ref. \cite{Neufeld:2008hs}.  There, the fast parton was treated as the source of an external color field interacting with a perturbative QGP through a Vlasov equation.  For a gluonic medium at temperature $T$ in the presence of a parton moving with velocity $\bald{u} = u \hat{z}$ at position $\bald{r} = u t\hat{z}$ in the relativistic limit ($\gamma = (1-u^2)^{-1/2} \gg 1$), the source is given by
\begin{equation}\label{compact}
J^\nu(x) = \left(J^0(x),{\bf u} J^0(x) - {\bf J}_\text{v}\right)
\end{equation}
where
\begin{eqnarray}\label{jnot}
J^0(\rho,z,t) &=& d(\rho,z,t) \gamma u^2 \left( 1 - \frac{\gamma u z_-}{\sqrt{z_-^2 \gamma^2 + \rho^2}} \right) \\
{\bf J}_\text{v}(\rho,z,t) &=& \left(\bald{x} - \bald{u} t \right) d(\rho,z,t) \frac{u^4}{\sqrt{z_-^2\gamma^2 + \rho^2}}
\end{eqnarray}
and
\begin{equation}
d(\rho,z,t) = \frac{\alpha_s ({Q_p^a})^2 m_{\rm D}^2}{8\pi(\rho^2 + \gamma^2 z_-^2)^{3/2}}.
\end{equation}
In the above expressions, $({Q_p^a})^2 = 3$ for a gluon and 4/3 for a quark, $\rho = (x^2 + y^2)^{1/2}$ is the radius transverse to the $z$ axis, $\alpha_s = g^2/4\pi$ is the strong coupling, $m_{\rm D} = gT$ and $z_- = (z - u t)$.  In what follows any numerical coefficient suppressed by powers of $\gamma^2$ will be dropped.  For instance, terms such as $\gamma^2 + 1$ will be taken as $\gamma^2$.

The vector part of the source, $\bald{J} = {\bf u} J^0 - {\bf J}_\text{v}$, is explicitly
\begin{equation}
\begin{split}\label{clumsyv}
\bald{J} &= \frac{\alpha_s ({Q_p^a})^2 m_{\rm D}^2 u^2}{8\pi}\left(\frac{\gamma\bald{u}}{(z_-^2 \gamma^2 + \rho^2)^{3/2}} - \frac{u^2\left(x,y,z_-\gamma^2\right)}{(z_-^2 \gamma^2 + \rho^2)^2}\right)
\end{split}
\end{equation}
which, as one can verify, can be re-written in the form of (\ref{generalj}),
\begin{equation}\label{jv}
\begin{split}
\bald{J} &= \frac{\alpha_s ({Q_p^a})^2 m_{\rm D}^2 u^2}{8\pi}\left(\frac{\gamma\bald{u}}{(z_-^2 \gamma^2 + \rho^2)^{3/2}} + \bald{\nabla}\frac{u^2}{2 (z_-^2 \gamma^2 + \rho^2)}\right).
\end{split}
\end{equation}

\section{Delta Function Expansion and Momentum Space Representation of the Source}\label{deltamom}

At distances increasingly far from the fast parton, one expects that the source term will begin to look like a $\delta$ function.  Since hydrodynamics is a long distance effective theory, the hydrodynamic solutions (in the range of validity) should be dominated by the lowest order terms in an expansion of gradients of a $\delta$ function centered at the location of the source parton (a detailed discussion of this is given in \cite{Chesler:2007sv}).  Higher order terms in the expansion, which are sensitive to the detailed structure of the source term, will become less important at larger distances.  In this section, I will expand the source term, as given by (\ref{jnot}) and (\ref{jv}), up to first order in gradients of a $\delta$ function.  Later, the hydrodynamic equations, (\ref{eps}), (\ref{gl}), (\ref{gt}), will be solved for both the full source term and the truncated series.  A comparison of the solutions will highlight at what distance scales the detailed structure of the source term becomes negligible.  It will prove convenient to Fourier transform the source into momentum space, which I will also do in this section, before attempting to solve the hydrodynamic equations.  The effect of color screening, which is absent in (\ref{jnot}) and (\ref{jv}), will be modelled by including a damping factor of the form $e^{-\rho \,{\rm m}}$, where ${\rm m}^{-1}$ is a typical screening scale.  In a perturbative QGP the inverse screening scale is given by $m_{\rm D} = gT$, which appears as a coefficient in front of the source term used here.  However, in principle, at higher orders the screening scale may be different than what appears as the coefficient of the source term.  It is thus instructive to keep ${\rm m}$ arbitrary; however, in solving the hydrodynamical equations in Sec.\ref{solving} I will set ${\rm m} = g T$.  Also, when necessary, a short distance cutoff will be used to regulate ultraviolet divergences.  A common choice, which will also be used here, for the short distance cutoff in collisional energy loss is $\rho_{min} = (2 \sqrt{E_p T})^{-1}$, where $E_p$ is the energy of the fast parton (see, for instance, \cite{Thoma:1991ea}).

Consider (\ref{jnot}), which can be expanded as
\begin{equation}
\begin{split}\label{deltajnot}
J^0(\rho,z,t) = C_0 \delta(\bald{x}_-) + \bald{C}_1 \cdot \bald{\nabla} \delta(\bald{x}_-) + \dots
\end{split}
\end{equation}
where I have used the shorthand notation
\begin{equation}
\delta(\bald{x}_-) \equiv \delta(x)\delta(y)\delta(z - u t).
\end{equation}
The coefficients, $C_0$ and $\bald{C}_1$, are found by taking the appropriate moment of $J^0(\rho,z,t)$.  Introducing the damping factor, $e^{-\rho \,{\rm m}}$, one has for $C_0$,
\begin{equation}
\begin{split}\label{czjnot}
C_0 &= \int d^3 x \, J^0(\rho,z,t) \, e^{-\rho \,{\rm m}} \\
&= \int d^3 x \frac{\alpha_s ({Q_p^a})^2 m_{\rm D}^2 \gamma u^2}{8\pi(\rho^2 + \gamma^2 z_-^2)^{3/2}} e^{-\rho \,{\rm m}} \\
&= \frac{\alpha_s ({Q_p^a})^2 m_{\rm D}^2 u^2}{2} \, G_0\left(\frac{{\rm m}}{2 \sqrt{E_p T}}\right) ,
\end{split}
\end{equation}
where $(2 \sqrt{E_p T})^{-1}$ has been introduced as a short distance cutoff, and $G_0$ is a representation of the incomplete Gamma function
\begin{equation}\label{incgamma}
G_0(z) = \int_{z}^{\infty} dt \frac{e^{-t}}{t}.
\end{equation}
The coefficient given by (\ref{czjnot}) gives the total energy deposited into the medium per unit time.

Similarly, $\bald{C}_1$ can be obtained as
\begin{equation}
\begin{split}\label{c1jnot}
\bald{C}_1 &= - \int d^3 x \, (x,y,z_-) \, J^0(\rho,z,t) e^{-\rho \,{\rm m}} \\
&= \int d^3 x \, \frac{\alpha_s ({Q_p^a})^2 m_{\rm D}^2 \gamma^2 u^3}{8\pi} \frac{z_- \, (x,y,z_-)}{(\rho^2 + \gamma^2 z_-^2)^{2}} e^{-\rho \,{\rm m}} \\
&= \frac{\alpha_s ({Q_p^a})^2 m_{\rm D}^2 u^2}{2} \left(0,0,\frac{\pi u }{4\, {\rm m} \gamma}\right).
\end{split}
\end{equation}
The results from (\ref{czjnot}) and (\ref{c1jnot}), together with (\ref{deltajnot}), give
\begin{equation}
\begin{split}\label{jnotD}
J_D^0(\rho,z,t) &= \frac{\alpha_s ({Q_p^a})^2 m_{\rm D}^2 u^2}{2} \\
&\times \left(\, G_0\left(\frac{{\rm m}}{2 \sqrt{E_p T}}\right) + \frac{\pi u }{4 \, {\rm m} \gamma} \partial_z \right)\delta(\bald{x}_-) + \dots
\end{split}
\end{equation}
where the subscript $_D$ is meant to indicate the expansion in gradients of a $\delta$ function.  Proceeding in an analogous manner yields for (\ref{jv})
\begin{equation}
\begin{split}\label{jvD}
\bald{J}_D&(\rho,z,t) = \frac{\alpha_s ({Q_p^a})^2 m_{\rm D}^2 u^2}{2} \left(\bald{u} \, G_0\left(\frac{{\rm m}}{2 \sqrt{E_p T}}\right) \right.\\
&\left.+ \frac{\pi}{8 \, {\rm m} \gamma} \left(u^2\bald{\nabla} +\bald{u}(\bald{u}\cdot\bald{\nabla})\right)\right)\delta(\bald{x}_-) + \dots
\end{split}
\end{equation}
Equations (\ref{jnotD}) and (\ref{jvD}) provide the expansion of the full source, (\ref{jnot}) and (\ref{jv}), up to first order in gradients of a $\delta$ function.  

As previously mentioned, it is easiest to solve for the hydrodynamics in Fourier space.  To do this, it is necessary to first transform the source terms into momentum space, following the general rule (\ref{generalrule}).  The details of the Fourier transforms of (\ref{jnot}) and (\ref{jv}) are give in Appendix A, and the result is
\begin{widetext}
\begin{equation}\label{fourfullsource}
\begin{split}
J^0(\bald{k},\omega) &= \frac{\alpha_s ({Q_p^a})^2 m_{\rm D}^2 u^2}{2} (2 \pi) \delta(\omega - u k_z) \left(\, G_0\left(\frac{{\rm m} + k_T}{2 \sqrt{E_p T}}\right) + \frac{i \pi (\bald{u}\cdot \bald{k})}{4 \gamma \sqrt{k_T^2 + {\rm m}^2}} \right) \\
\bald{J}(\bald{k},\omega) &= \frac{\alpha_s ({Q_p^a})^2 m_{\rm D}^2 u^2}{2} (2 \pi) \delta(\omega - u k_z) \\
&\times \left(\bald{u}\, G_0\left(\frac{{\rm m} + k_T}{2 \sqrt{E_p T}}\right) + \frac{i \pi}{4 \gamma}\left(\frac{\sqrt{k_T^2 + {\rm m}^2} - {\rm m}}{k_T^2}\right)\left(u^2\bald{k} +\frac{\bald{u}(\bald{u}\cdot\bald{k}){\rm m}}{\sqrt{k_T^2 + {\rm m}^2}}\right)\right)
\end{split}
\end{equation}
These equations should be compared to the Fourier transforms of (\ref{jnotD}) and (\ref{jvD}), which are found by making the replacements $\delta(z_-)\rightarrow (2 \pi) \delta(\omega - u k_z)$ and $\bald{\nabla} \rightarrow i\bald{k}$:
\begin{equation}
\begin{split}\label{fourDsource}
J_D^0(\bald{k},\omega) &= \frac{\alpha_s ({Q_p^a})^2 m_{\rm D}^2 u^2}{2} (2 \pi) \delta(\omega - u k_z) \left(\, G_0\left(\frac{{\rm m} }{2 \sqrt{E_p T}}\right) + \frac{i \pi (\bald{u}\cdot \bald{k})}{4 \,{\rm m} \gamma} \right) \\
\bald{J}_D(\bald{k},\omega) &= \frac{\alpha_s ({Q_p^a})^2 m_{\rm D}^2 u^2}{2} (2 \pi) \delta(\omega - u k_z) \left(\bald{u} \, G_0\left(\frac{{\rm m}}{2 \sqrt{E_p T}}\right) + \frac{i \pi}{8 \, {\rm m}\gamma} \left(u^2\bald{k} +\bald{u}(\bald{u}\cdot\bald{k})\right)\right)
\end{split}
\end{equation}
\end{widetext}
One can verify that (\ref{fourfullsource}) reduces to (\ref{fourDsource}) by taking the $k_T \rightarrow 0$ limit in the coefficients of $1$ and $\bald{k}$.  (\ref{fourfullsource}) and (\ref{fourDsource}) will be used in the next section to solve for the hydrodynamic variables of the medium.

As discussed at the beginning of the section, at distances increasingly far from the fast parton the source term is dominated by the lowest order terms in an expansion of gradients of a $\delta$ function.  It's clear from inspection that the detailed structure of the full source term, given by (\ref{fourfullsource}), becomes important at a momentum scale $k \sim {\rm m}$.  This could have been anticipated, since the full source term is calculated up to distances of the order of the screening length.  A quantitative comparison of the effects of (\ref{fourfullsource}) and (\ref{fourDsource}) requires solving the hydrodynamic equations.  This will be done in the next section.

\section{Solving the Equations}\label{solving}

The result given by (\ref{fourfullsource}) is combined with equations (\ref{eps}), (\ref{gl}), and (\ref{gt}) to yield $\delta\epsilon ({\mathbf k},\omega)$, ${\mathbf g}_L ({\mathbf k},\omega)$ and ${\mathbf g}_T ({\mathbf k},\omega)$.  These are then transformed back to position space using the relation (\ref{generalrule}).  In $\delta\epsilon ({\mathbf k},\omega)$ and ${\mathbf g}_L ({\mathbf k},\omega)$, one can find $J_L$ by taking $\hat{\mathbf k}\cdot \bald{J}$, which is written conveniently as
\begin{widetext}
\begin{equation}
\begin{split}
J_L &= \frac{\alpha_s ({Q_p^a})^2 m_{\rm D}^2 u^2}{2 k} (2 \pi) \delta(\omega - u k_z) \left((\bald{u}\cdot\bald{k})\left(\, G_0\left(\frac{m_D + k_T}{2 \sqrt{E_p T}}\right) + \frac{i \pi (\bald{u}\cdot\bald{k})}{4 \gamma \sqrt{k_T^2 + m_{\rm D}^2}}\right)+ \frac{i \pi u^2}{4 \gamma}\left(\sqrt{k_T^2 + m_{\rm D}^2} - m_{\rm D}\right)\right) \\
&= \frac{(\bald{u}\cdot\bald{k}) J^0}{k} + \frac{\alpha_s ({Q_p^a})^2 m_{\rm D}^2 u^2}{2 k} (2 \pi) \delta(\omega - u k_z) \frac{i \pi u^2}{4 \gamma}\left(\sqrt{k_T^2 + m_{\rm D}^2} - m_{\rm D}\right)
\end{split}
\end{equation}
where I am now taking ${\rm m} = m_{\rm D}$.  After integrating out $\delta(\omega - u k_z)$ and using (\ref{bessel}), the expression for $\delta\epsilon ({\mathbf x}, t)$ is given by
\begin{equation}
\begin{split}\label{contoureps}
\delta\epsilon({\bf x},t) &= \frac{\alpha_s ({Q_p^a})^2 m_{\rm D}^2 u^2 \lambda^2}{8 \pi^2 c_s^2}\int d k_T d k_z \frac{k_T J_0(\rho k_T)e^{i k_z (z - u t)}}{{k_z^2 - \lambda^2 k_T^2 + i \sigma}} \times \\
&\left(\left(\, G_0\left(\frac{m_D + k_T}{2 \sqrt{E_p T}}\right) + \frac{i \pi u k_z}{4 \gamma \sqrt{k_T^2 + m_{\rm D}^2}}\right)\left(2 i u k_z - \Gamma_s k^2\right) - \frac{\pi u^2}{4 \gamma}\left(\sqrt{k_T^2 + m_{\rm D}^2} - m_{\rm D}\right)\right)
\end{split}
\end{equation}
where $\lambda^2 = c_s^2/(u^2-c_s^2)$, $\sigma = \Gamma_s u (\lambda^2/c_s^2) k_z(k_T^2 + k_z^2)$ and I am again working in plane polar coordinates.  The integral over $k_z$ can be performed using contour integration.  Poles are located at $k_z = \pm (k_T^2 \lambda^2 \mp i |\sigma|)^{1/2}$, where $|\sigma|$ is itself a function of $k_z$.  When evaluating the residues at these poles I make the approximation $\sigma(k_z) \approx \sigma(\pm k_T \lambda)$.  This approximation is valid at momentum scales for which the sound attenuation is small ($k_T \ll c_s^2/\Gamma_s$), which should be reasonable in the hydrodynamic limit.  Both poles are located in the lower complex plane so that the integration only contributes for $z < u t$, i.~e., behind the source parton.  Performing the integration yields
\begin{equation}
\begin{split}\label{finaleps}
\delta\epsilon({\bf x},t) &= \frac{\alpha_s ({Q_p^a})^2 m_{\rm D}^2 u^2 \lambda^2}{4 \pi c_s^2}\text{Re}\left[\int d k_T \frac{i J_0(\rho k_T)e^{i k_T \sqrt{\lambda^2 - i k_T \alpha} (z - u t)}}{\sqrt{\lambda^2 - i k_T \alpha}} \left(\frac{\pi u^2}{4 \gamma}\left(\sqrt{k_T^2 + m_{\rm D}^2} - m_{\rm D}\right) \right. \right. \\
&\left.\left. - \left(G_0\left(\frac{m_D + k_T}{2 \sqrt{E_p T}}\right) + \frac{i \pi u k_T \sqrt{\lambda^2 - i k_T \alpha}}{4 \gamma \sqrt{k_T^2 + m_{\rm D}^2}}\right)\left(2 i u k_T \sqrt{\lambda^2 - i k_T \alpha} - \Gamma_s k_T^2(1 + \lambda^2 - i k_T \alpha)\right)\right)\right]
\end{split}
\end{equation}
where $\alpha \equiv \Gamma_s u \lambda^3/(c_s^2(\lambda^2 + 1))$.  The final integration over $k_T$ is performed numerically.  The analogous expression resulting from (\ref{fourDsource}) is given by
\begin{equation}
\begin{split}\label{finalepsD}
{\delta\epsilon}_D({\bf x},t) &= \frac{\alpha_s ({Q_p^a})^2 m_{\rm D}^2 u^2 \lambda^2}{4 \pi c_s^2}\text{Re}\left[\int d k_T \frac{i J_0(\rho k_T)e^{i k_T \sqrt{\lambda^2 - i k_T \alpha} (z - u t)}}{\sqrt{\lambda^2 - i k_T \alpha}} \left(\frac{\pi u^2 k_T^2}{8 \gamma m_{\rm D}} \right. \right. \\
&\left.\left. - \left(G_0\left(\frac{m_D}{2 \sqrt{E_p T}}\right) + \frac{i \pi u k_T\sqrt{\lambda^2 - i k_T \alpha}}{4 \gamma m_{\rm D}}\right)\left(2 i u k_T \sqrt{\lambda^2 - i k_T \alpha} - \Gamma_s k_T^2(1 + \lambda^2 - i k_T \alpha)\right)\right)\right]
\end{split}
\end{equation}

The same approach is applied to ${\mathbf g}_L ({\mathbf x}, t)$.  The contour integration proceeds in the same manner as in (\ref{contoureps}) with the exception that one term has additional poles at $k_z = \pm i k_T$.  The additional pole at $k_z = i k_T$ allows for some contribution in the region in front of the source parton.  The result is
\begin{equation}
\begin{split}\label{glresult}
&\text{for } z <  u t:\\
{\mathbf g}_L& ({\mathbf x}, t) = \frac{\alpha_s ({Q_p^a})^2 m_{\rm D}^2 u^2 \lambda^2}{4 \pi c_s^2}\text{Re}\left[\int d k_T \frac{k_T \, e^{i k_T\sqrt{\lambda^2 - i k_T \alpha}(z - u t)}}{\sqrt{\lambda^2 - i k_T \alpha}} \left(\frac{x}{\rho} i J_1(\rho k_T), \frac{y}{\rho} i J_1(\rho k_T),\sqrt{\lambda^2 - i k_T \alpha} J_0(\rho k_T)\right)\times \right. \\
&\left. \left(G_0\left(\frac{m_D + k_T}{2 \sqrt{E_p T}}\right)\left(\frac{u^2 (\lambda^2 - i k_T \alpha)}{(1 + \lambda^2 - i k_T \alpha)} + c_s^2\right) + \frac{i \pi u k_T \,\sqrt{\lambda^2 - i k_T \alpha}}{4 \gamma \,\sqrt{k_T^2 + m_{\rm D}^2} } \left((u^2 + c_s^2) + \frac{u^2 m_D(m_{\rm D} - \sqrt{k_T^2 + m_{\rm D}^2})}{k_T^2 (1 + \lambda^2 - i k_T \alpha)}\right)\right)\right] \\
& + \frac{\alpha_s ({Q_p^a})^2 m_{\rm D}^2 u^2}{8 \pi } \int d k_T \, e^{- k_T |z - u t|} \left(-\frac{x}{\rho} J_1(\rho k_T) , - \frac{y}{\rho} J_1(\rho k_T) , J_0(\rho k_T)\right) \\
& \times \left(k_T \, G_0\left(\frac{m_D + k_T}{2 \sqrt{E_p T}}\right) + \frac{\pi m_D}{4 \gamma} \left(1 - \frac{m_{\rm D}}{\sqrt{k_T^2 + m_{\rm D}^2}}\right)\right) \\
&\text{for } z > u t:\\
{\mathbf g}_L& ({\mathbf x}, t) = -\frac{\alpha_s ({Q_p^a})^2 m_{\rm D}^2 u^2}{8 \pi } \int d k_T \, e^{- k_T |z - u t|} \left(\frac{x}{\rho} J_1(\rho k_T) , \frac{y}{\rho} J_1(\rho k_T) , J_0(\rho k_T)\right) \\
& \times \left(k_T \, G_0\left(\frac{m_D + k_T}{2 \sqrt{E_p T}}\right) - \frac{\pi m_D}{4 \gamma} \left(1 - \frac{m_{\rm D}}{\sqrt{k_T^2 + m_{\rm D}^2}}\right)\right).
\end{split}
\end{equation}
which must be done numerically.  The analogous expression resulting from (\ref{fourDsource}) is given by
\begin{equation}
\begin{split}\label{glresultD}
&\text{for } z <  u t:\\
{{\mathbf g}_L}_D& ({\mathbf x}, t) = \frac{\alpha_s ({Q_p^a})^2 m_{\rm D}^2 u^2 \lambda^2}{4 \pi c_s^2}\text{Re}\left[\int d k_T \frac{k_T \, e^{i k_T\sqrt{\lambda^2 - i k_T \alpha}(z - u t)}}{\sqrt{\lambda^2 - i k_T \alpha}} \left(\frac{x}{\rho} i J_1(\rho k_T), \frac{y}{\rho} i J_1(\rho k_T),\sqrt{\lambda^2 - i k_T \alpha} J_0(\rho k_T)\right)\times \right. \\
&\left. \left(G_0\left(\frac{m_D}{2 \sqrt{E_p T}}\right)\left(\frac{u^2 (\lambda^2 - i k_T \alpha)}{(1 + \lambda^2 - i k_T \alpha)} + c_s^2\right) + \frac{i \pi u k_T \,\sqrt{\lambda^2 - i k_T \alpha}}{4 \gamma \,m_{\rm D}} \left((u^2 + c_s^2) + \frac{u^2}{2(1 + \lambda^2 - i k_T \alpha)}\right)\right)\right] \\
& + \frac{\alpha_s ({Q_p^a})^2 m_{\rm D}^2 u^2}{8 \pi } \int d k_T \, e^{- k_T |z - u t|} \left(-\frac{x}{\rho} J_1(\rho k_T) , - \frac{y}{\rho} J_1(\rho k_T) , J_0(\rho k_T)\right)\left(k_T \, G_0\left(\frac{m_D}{2 \sqrt{E_p T}}\right) + \frac{\pi k_T^2}{8 \gamma \, m_D}\right) \\
&\text{for } z > u t:\\
{{\mathbf g}_L}_D & ({\mathbf x}, t) = -\frac{\alpha_s ({Q_p^a})^2 m_{\rm D}^2 u^2}{8 \pi } \int d k_T \, e^{- k_T |z - u t|} \left(\frac{x}{\rho} J_1(\rho k_T) , \frac{y}{\rho} J_1(\rho k_T) , J_0(\rho k_T)\right)\left(k_T \, G_0\left(\frac{m_D}{2 \sqrt{E_p T}}\right) - \frac{\pi k_T^2}{8 \gamma \, m_D}\right).
\end{split}
\end{equation}

Combining (\ref{jt}) with (\ref{gt}) and following the same approach used above allows for the determination of ${\mathbf g}_T ({\mathbf x}, t)$.  The denominator of (\ref{gt}) has a simpler structure than in (\ref{eps},\ref{gl}) which allows the contour integration to be performed exactly.
Defining $\Omega \equiv 4 u/3 \Gamma_s$ it is found that
\begin{equation}
\begin{split}\label{gtresult}
\bald{g}_T({\mathbf x}, t) &= \frac{\alpha_s ({Q_p^a})^2 m_{\rm D}^2}{4 \pi }\int d k_T \frac{k_T^2 \, e^{\pm\varDelta^{\mp}|z - u t|}}{(k_T^2 - (\varDelta^{\mp})^2)\sqrt{1 + \frac{4 k_T^2}{\Omega^2}}} \times \\
&\left(\, G_0\left(\frac{m_D + k_T}{2 \sqrt{E_p T}}\right) - \frac{\pi \, m_{\rm D} \varDelta^{\mp}}{4 \gamma}\left(\frac{\sqrt{k_T^2 + m_{\rm D}^2} - m_{\rm D}}{k_T^2 \sqrt{k_T^2 + m_{\rm D}^2}}\right)\right)\left(J_1(\rho k_T)\frac{x}{\rho} \varDelta^{\mp}, J_1(\rho k_T)\frac{y}{\rho}\varDelta^{\mp}, J_0(\rho k_T) k_T\right) \\
& \pm  \frac{\alpha_s ({Q_p^a})^2 m_{\rm D}^2}{16 u \pi }\int d k_T \, e^{- k_T |z -u t|} \left(\pm J_1(\rho k_T)\frac{x}{\rho} k_T, \pm J_1(\rho k_T)\frac{y}{\rho} k_T, - J_0(\rho k_T) k_T \right) \\
&\times \left(\, G_0\left(\frac{m_D + k_T}{2 \sqrt{E_p T}}\right) \mp \frac{\pi \, m_{\rm D} k_T}{4 \gamma}\left(\frac{\sqrt{k_T^2 + m_{\rm D}^2} - m_{\rm D}}{k_T^2 \sqrt{k_T^2 + m_{\rm D}^2}}\right)\right)
\end{split}
\end{equation}
where $\mp$ refers to the sign of $(z - u t)$ and
\begin{equation}
\varDelta^{\mp} \equiv \frac{\Omega}{2}\left(1 \mp \sqrt{1 + \frac{4 k_T^2}{\Omega^2}}\right).
\end{equation}
The analogous result from (\ref{fourDsource}) is given by
\begin{equation}
\begin{split}\label{gtresultD}
{\bald{g}_T}_D&({\mathbf x}, t) = \frac{\alpha_s ({Q_p^a})^2 m_{\rm D}^2}{4 \pi }\int d k_T \frac{k_T^2 \, e^{\pm\varDelta^{\mp}|z - u t|}}{(k_T^2 - (\varDelta^{\mp})^2)\sqrt{1 + \frac{4 k_T^2}{\Omega^2}}} \times \\
&\left(\, G_0\left(\frac{m_D}{2 \sqrt{E_p T}}\right) - \frac{\pi \, \varDelta^{\mp}}{8 \gamma m_{\rm D}}\right)\left(J_1(\rho k_T)\frac{x}{\rho} \varDelta^{\mp}, J_1(\rho k_T)\frac{y}{\rho}\varDelta^{\mp}, J_0(\rho k_T) k_T\right) \\
& \pm  \frac{\alpha_s ({Q_p^a})^2 m_{\rm D}^2}{16 u \pi }\int d k_T \, e^{- k_T |z -u t|} \left(\pm J_1(\rho k_T)\frac{x}{\rho} k_T, \pm J_1(\rho k_T)\frac{y}{\rho} k_T, - J_0(\rho k_T) k_T \right)\left(\, G_0\left(\frac{m_D}{2 \sqrt{E_p T}}\right) \mp \frac{\pi \,k_T}{8 \gamma m_{\rm D}}\right).
\end{split}
\end{equation}
\end{widetext}

\begin{figure*}
\centerline{
\includegraphics[width = 0.85\linewidth]{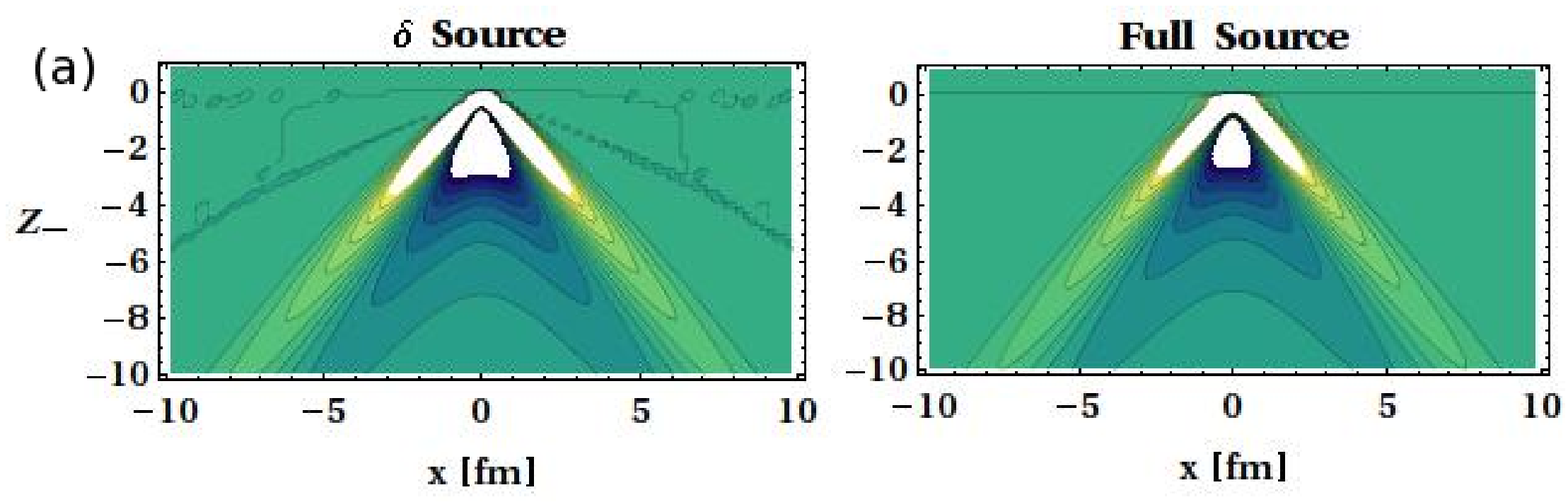}
}
\centerline{
\includegraphics[width = 0.85\linewidth]{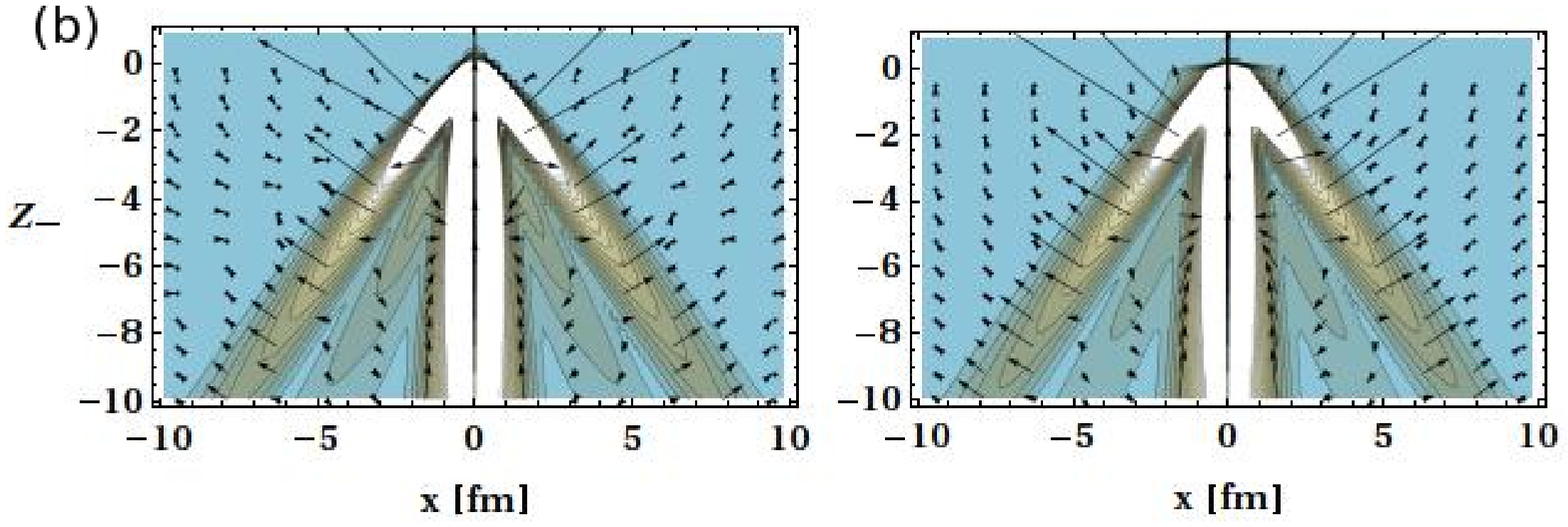}
}
\caption{(Color online) Plots of (a): the perturbed energy density, and (b): the perturbed momentum density, for both the full source term and the $\delta$ function expanded source term.  Here, $\eta/s = 1/4\pi$ and $c_s = 0.57$.  The direction of the momentum density is indicated by the arrows. 
}
\label{contourdelta}
\end{figure*}

\begin{figure*}
\centerline{
\includegraphics[width = 0.5\linewidth]{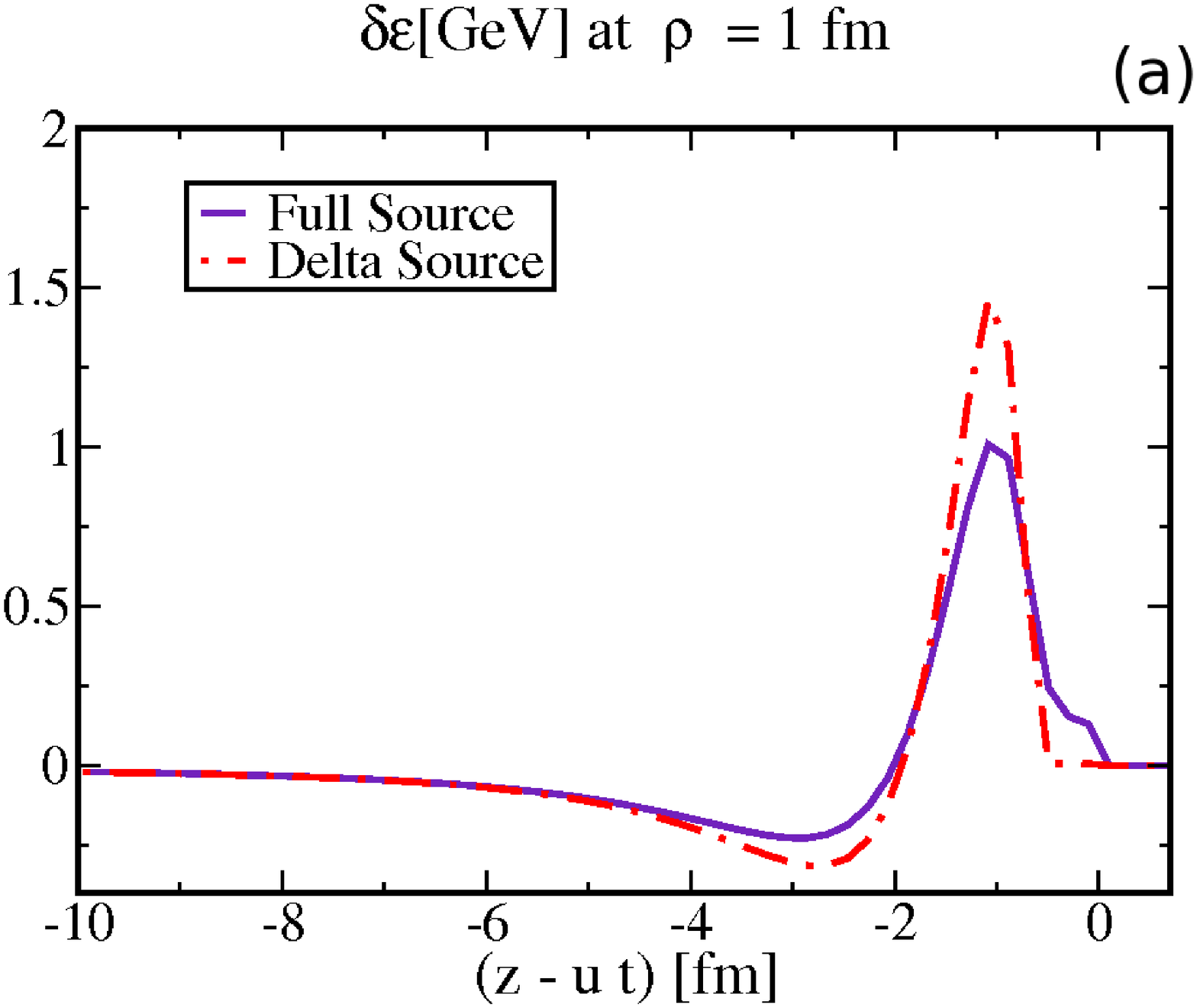}
\includegraphics[width = 0.5\linewidth]{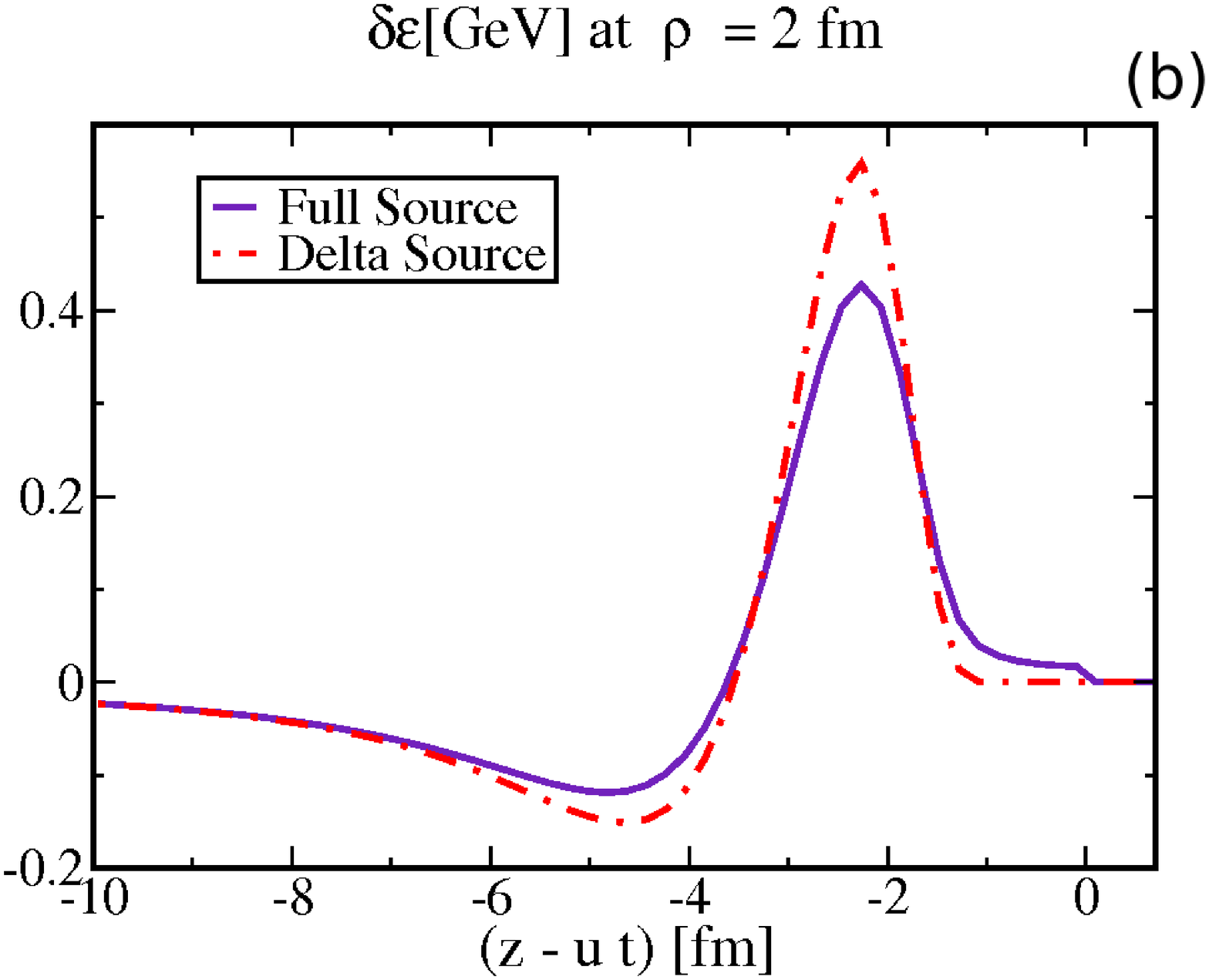}
}
\caption{(Color online) A comparison of the perturbed energy density generated by the full source term and the $\delta$ function expanded source term for two different values of $\rho$.  The two results converge at a distance of about 5-6 fm behind the source parton for the chosen values of $\rho$.
}
\label{lineardelta}
\end{figure*}

\begin{figure*}
\centerline{
\includegraphics[width = 1.0\linewidth]{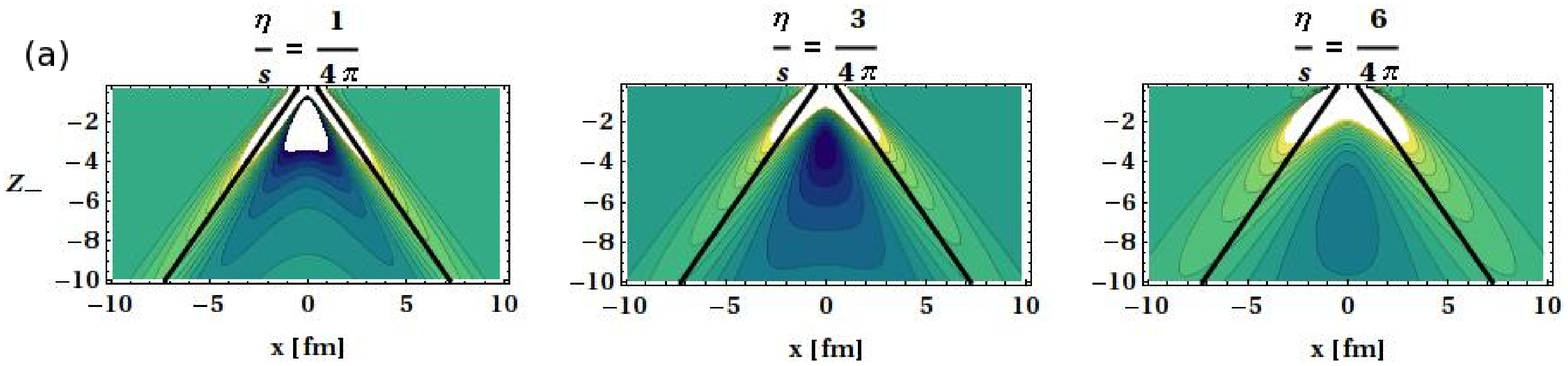}
}
\centerline{
\includegraphics[width = 1.0\linewidth]{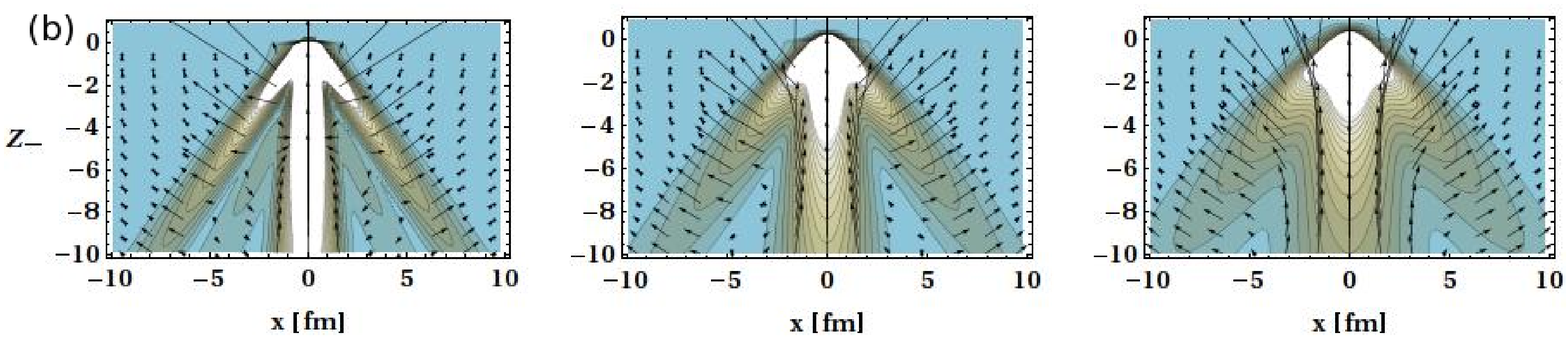}
}
\caption{(Color online) Plots of (a): the perturbed energy density, and (b): the perturbed momentum density for different values of the shear viscosity to entropy density ratio, $\eta/s$.  The black lines in (a) are drawn where one would expect to find the boundary of a Mach cone in the absence of dissipative effects.  Plots scaled by the radius, $\rho$, which factor in the conical broadening of the cone, are shown in Figure \ref{3dvisc}.}
\label{contourvisc}
\end{figure*}

\begin{figure*}
\centerline{
\includegraphics[width = 1.0\linewidth]{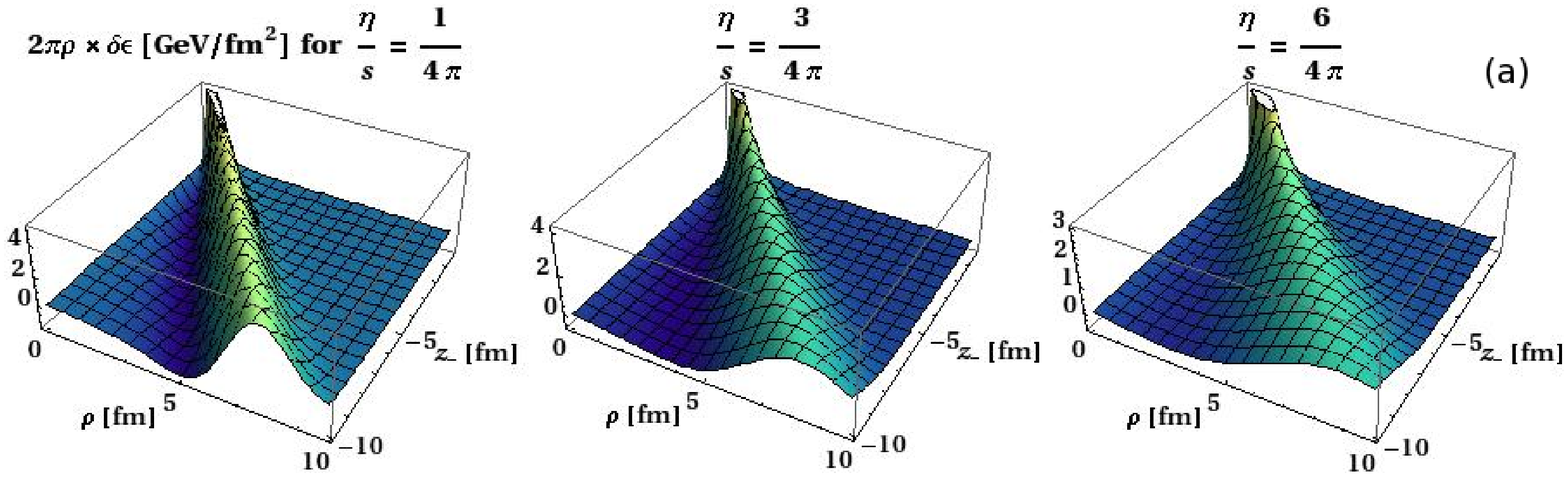}
}
\centerline{
\includegraphics[width = 1.0\linewidth]{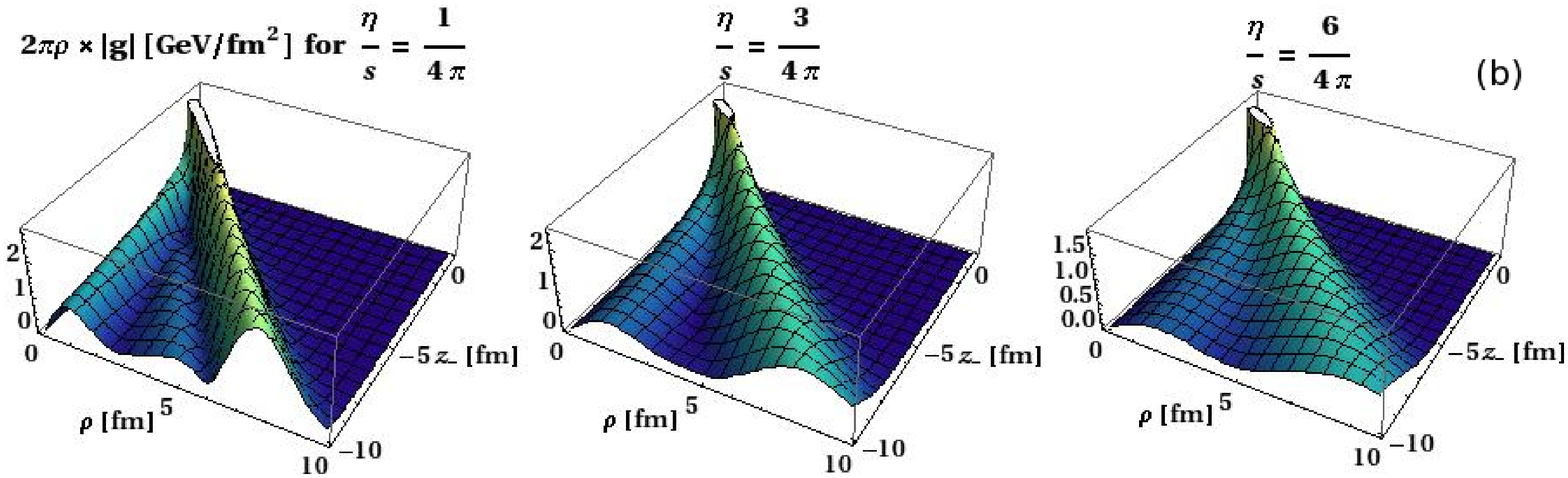}
}
\caption{(Color online) Plots of (a): the total perturbed energy density, and (b): the total perturbed momentum density, contained at a given radius in the $\rho-z_-$ plane for different values of $\eta/s$.  As one can see in (b) the total perturbed momentum density carried by the sonic Mach cone exceeds that contained in the diffusive wake.}
\label{3dvisc}
\end{figure*}

\begin{figure*}
\centerline{
\includegraphics[width = 1.0\linewidth]{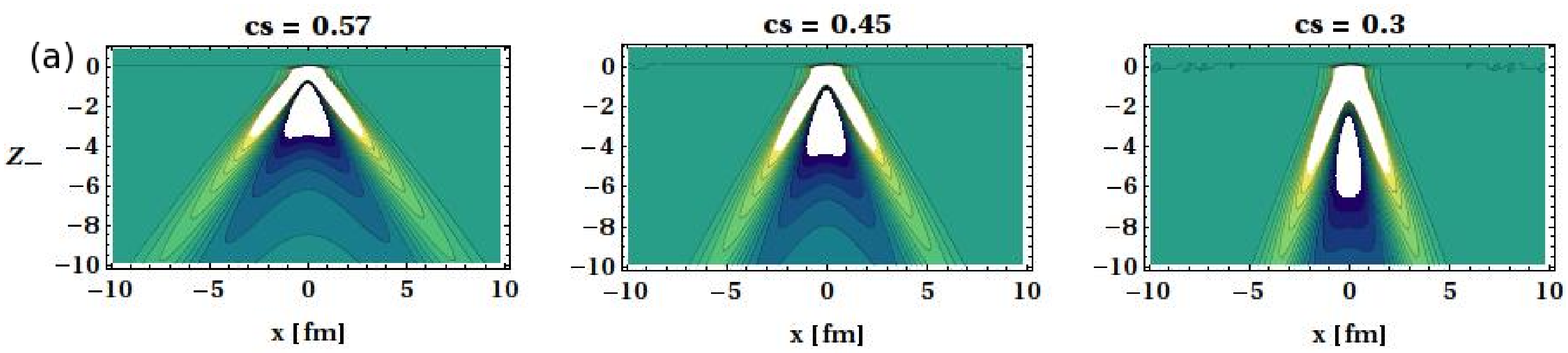}
}
\centerline{
\includegraphics[width = 1.0\linewidth]{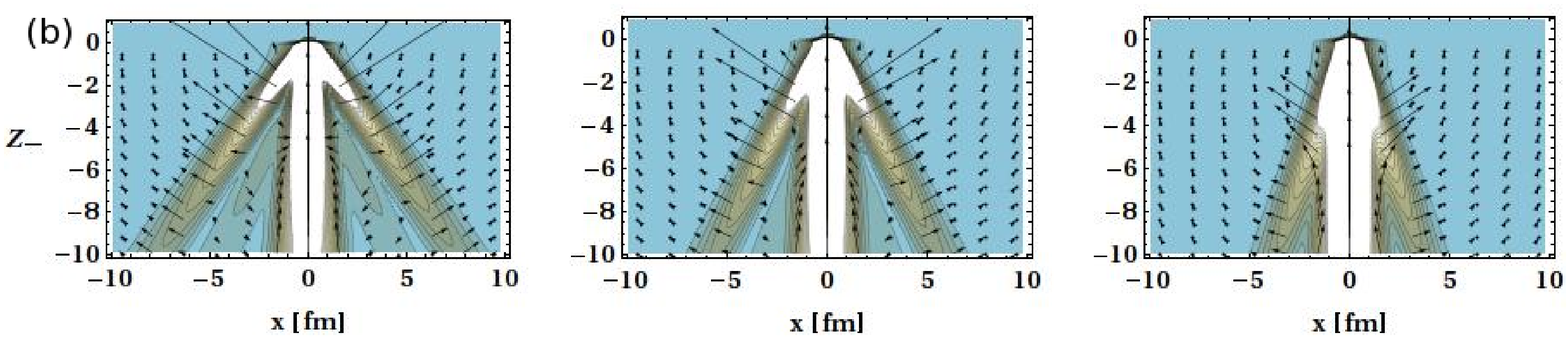}
}
\caption{(Color online) Plots of (a): the perturbed energy density, and (b): the perturbed momentum density for different values of the speed of sound, $c_s$.  Plots scaled by the radius, $\rho$, which factor in the conical broadening of the cone, are shown in Figure \ref{3dsound}.}
\label{contoursound}
\end{figure*}

\begin{figure*}
\centerline{
\includegraphics[width = 1.0\linewidth]{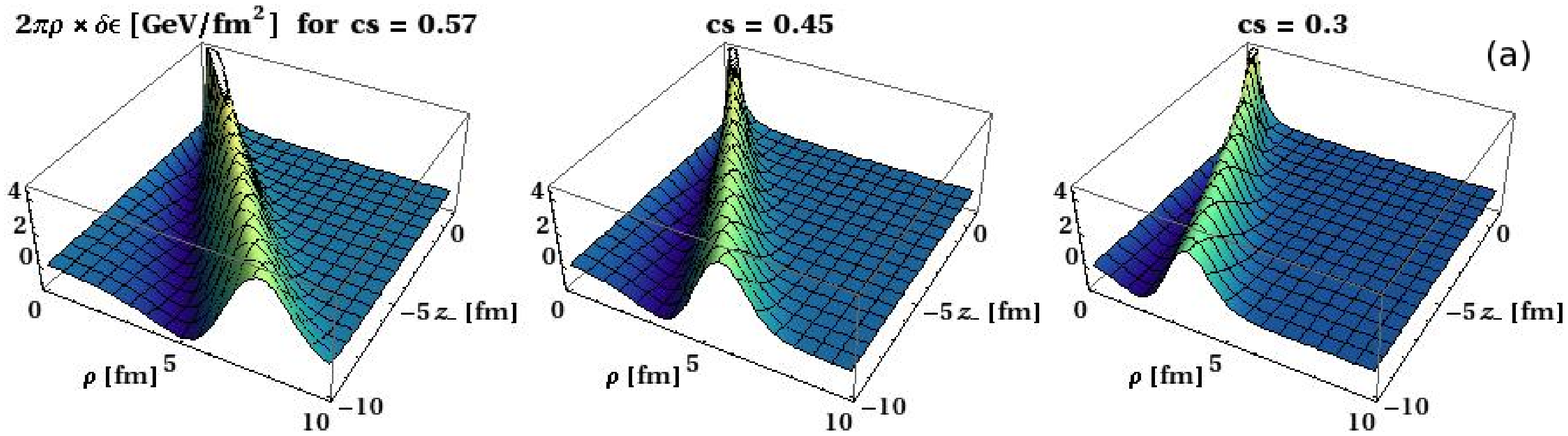}
}
\centerline{
\includegraphics[width = 1.0\linewidth]{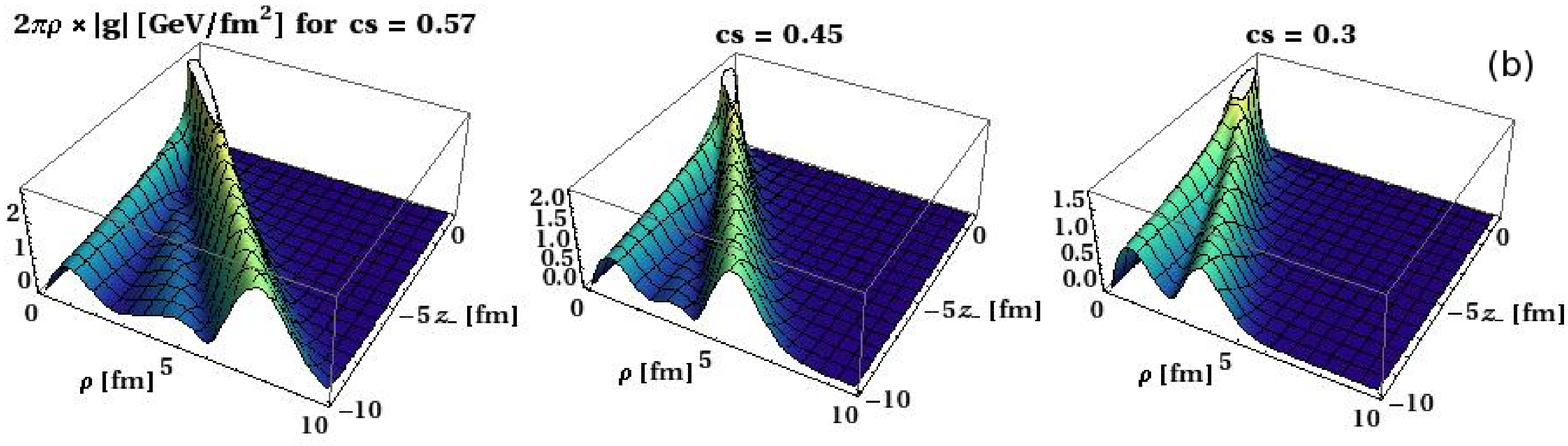}
}
\caption{(Color online) Plots of (a): the total perturbed energy density, and (b): the total perturbed momentum density, contained at a given radius in the $\rho-z_-$ plane for different values of $c_s$.  The magnitude of the Mach cone, when integrated, is similar for the different plots.}
\label{3dsound}
\end{figure*}

\section{Numerical Results and Discussion}\label{randd}

Having obtained expressions for the hydrodynamic quantities $\delta\epsilon({\bf x},t)$, $\bald{g}_L({\bf x},t)$ and $\bald{g}_T({\bf x},t)$ I now consider the results of numerical integration.  All calculations are performed for a gluon moving along the positive $z$ axis at position $ut$ and speed $u = 0.99955 c \text{ }(\gamma \approx 33)$.  The strong coupling, $\alpha_s$, is chosen to be $1/\pi$, the temperature is taken to be $T = 350$ MeV, and $E_p = 16$ GeV.  As mentioned in the introduction, I will compare the solutions resulting from the full source term and the $\delta$ function expanded source term, given by (\ref{fourfullsource}) and (\ref{fourDsource}), respectively.  I will also make a comparison of the results for a range of values of the shear viscosity to entropy density ratio, $\eta/s$, and speed of sound, $c_s$.  

The solutions resulting from the the full source term and the $\delta$ function source are shown in Figures \ref{contourdelta} and \ref{lineardelta}.  In both Figures the results are plotted for $\eta/s = 1/4\pi$ and $c_s = c/\sqrt{3}$.  Figure \ref{contourdelta}(a) shows a contour plot of $\delta\epsilon({\bf x},t)$ for each source.  In both cases a well define Mach cone is visible in the trailing medium.  Figure \ref{contourdelta}(b) shows the result for the magnitude of the momentum density, $|\bald{g}| = |\bald{g}_L({\bf x},t) + \bald{g}_T({\bf x},t)|$.  One now sees both a sound contribution from $\bald{g}_L({\bf x},t)$, which excites a Mach cone, and a diffusive contribution from $\bald{g}_T({\bf x},t)$, which is excited in the region directly behind the source gluon.  The diffusive momentum density produces flow almost exclusively in the direction of the source parton's velocity, while the Mach cone generates flow outward and perpendicular to it's boundary, as indicated by the arrows.  One can see from Figure \ref{contourdelta} that the full source term and the $\delta$ function expanded source term provide qualitatively similar results, particularly in the region far from the source parton.  In the region near $z_- = 0$, the full source solution has a noticeably larger transverse extent than the corresponding $\delta$ source result.  A more quantitative comparison can be made by examining Figure \ref{lineardelta} where $\delta\epsilon({\bf x},t)$ is plotted as a function of $z_-$ for fixed $\rho$.  Here one sees that the two results converge at a distance of about 5-6 fm behind the source parton for the chosen $\rho$ values.  

Results are next presented for three different values of the shear viscosity to entropy density ratio, $\eta/s$.  The first value chosen for $\eta/s$ is $1/4\pi\approx 0.08$, which has been proposed \cite{Kovtun:2004de} as a universal lower bound for all relativistic quantum field theories and is calculated in the strongly coupled limit.  The other two values for $\eta/s$ are multiples of the previous value, $3/4\pi$ and $6/4\pi$, and are more consistent with the application of perturbation theory, which is the method used to calculate the source used in this paper.  For example, Arnold {\em et al.} \cite{Arnold:2003zc} found for the leading order result $\eta/s = 0.48$ for a gluonic plasma with $\alpha_s = 0.3$.  More recently, Xu and Greiner found $\eta/s = 0.13$ for a gluonic plasma with the same value of $\alpha_s$ by going beyond leading order in the diluteness of the medium \cite{Xu:2007ns}.  A small value of the shear viscosity, which is required by the RHIC data \cite{Romatschke:2007mq}, is not necessarily incompatible with perturbation theory, especially if the viscosity is lowered by anomalous contributions \cite{Asakawa:2006tc}.

The results for $\delta\epsilon({\bf x},t)$ and $|\bald{g}| = |\bald{g}_L({\bf x},t) + \bald{g}_T({\bf x},t)|$ for all three viscosities are shown in Figures \ref{contourvisc} and \ref{3dvisc}, where $c_s = c/\sqrt{3}$.  The black lines in the contour plots of Figure \ref{contourvisc}(a) are drawn along the slope $x = \pm\lambda (z - u t)$ which is where one would expect to find the boundary of a Mach cone in the absence of dissipative effects.  In Figure \ref{3dvisc}, the total energy density (figure (a)), and magnitude of momentum density (figure (b)), contained at a given radius are shown.  It is clear from the plots that the Mach cone broadens and weakens as the viscosity is increased.  

Finally, results are shown for three different values of the speed of sound, $c_s$.  The first value is $c_s = 0.57$, which is the the limiting value for a conformal ideal relativistic gas, while the other two values are $c_s = 0.45, 0.3$.  It's likely that the QGP produced at RHIC experiences a speed of sound close to all three of these values during its evolution \cite{Karsch:2006sf}.  The results are shown in Figures \ref{contoursound} and \ref{3dsound}, where I have chosen $\eta/s = 1/4\pi$.  One should note that the diffusive contribution, $\bald{g}_T({\bf x},t)$, is independent of the speed of sound.

It is interesting to consider how the results presented here compare to experimental data.  In the di-hadron correlation functions measured at RHIC there is a double peak structure in the back-jet (source parton) distribution which has been interpreted by some as the result of Mach cone generated flow.  In the spectrum presented here one indeed finds Mach cone generated flow but also finds a substantial diffusive flow, which seems to be missing from the RHIC data.  In order to make a comparison, however, one must consider that the matter created in heavy ion collisions at RHIC rapidly expands in contrast to the static background assumed here.  The diffusive momentum is deposited locally and is thus probably difficult to observe in an expanding medium.  On the other hand, the Mach cone propagates at the speed of sound, which is of the same order of magnitude as the expansion velocities in the matter produced at RHIC, and is likely more readily observed experimentally.  

Care should be taken when examining the azimuthal particle spectrum generated by a fast parton using an isochronous Cooper-Frye freeze-out from a static medium, such as in the work done by Betz {\it et al.} \cite{Betz:2008wy}.  In their paper, the authors compared the perturbative QCD based source term studied here with one derived in the strongly coupled Ads/CFT correspondence \cite{Chesler:2007sv}.  Their conclusion was that the anomalous azimuthal hadron correlations observed at RHIC are likely the result of flow generated by the non-equilibrium Neck zone in the Ads/CFT case, a contribution which does not obey Mach's law \cite{Noronha:2008un}.  In the isochronous Cooper-Frye freeze-out scenario the entire volume of matter is assumed to hadronize at the same time, independent of physical processes.  This is in contrast to the freeze-out in a heavy ion collision, which occurs as the result of an expanding and cooling medium.  The effect of an isochronous freeze-out scenario is that any cylindrically symmetric, or conical, contributions tend to be washed out (see the discussion in \cite{Neufeld:2008eg}).  Any rigorous comparison to experimental results will require incorporating a realistic source term in an expanding medium.

In summary, I have here presented in detail a method of solution for the linearized hydrodynamical equations of a QGP coupled to the source term generated by a fast parton.  The solution has been examined for different values of the shear viscosity to entropy density ratio and speed of sound.  Additionally, the relevance of finite source structure has been investigated by performing an expansion in gradients of a $\delta$ function centered at the location of the source parton.  Comparison of the medium response generated by the full source with that generated by the $\delta$ expanded one shows that the result is sensitive to the finite structure up to distances of several fm from the fast parton for the source examined here.

{\it Acknowledgments:} I thank Berndt M\"uller for many discussions and advice.  This work was supported in part by the U.~S.~Department of Energy under grant DE-FG02-05ER41367.

\section*{Appendix A: Obtaining the Fourier Representation of (\ref{jnot}) and (\ref{jv})}  
\renewcommand{\theequation}{A-\arabic{equation}}
  \setcounter{equation}{0}  
\renewcommand{\thefigure}{A-\arabic{figure}}
  \setcounter{figure}{0}  

In Sec.\ref{deltamom} the explicit determination of the Fourier transform of the full source was put off to this appendix.  Including the damping factor, $e^{-\rho \, {\rm m}}$, one has for the inverse Fourier transforms of (\ref{jnot}) and (\ref{jv})
\begin{equation}
\begin{split}\label{fourjvec}
\bald{J}&(\bald{k},\omega) = \frac{\alpha_s ({Q_p^a})^2 m_{\rm D}^2 u^2}{8\pi} \times\\
&\int d^4 x \, e^{i k \cdot x - \rho \, {\rm m}} \left(\frac{\gamma\,\bald{u}}{(z_-^2 \gamma^2 + \rho^2)^{3/2}} + \bald{\nabla}\frac{u^2}{2 (z_-^2 \gamma^2 + \rho^2)}\right)
\end{split}
\end{equation}
and
\begin{equation}
\begin{split}\label{fourjnot}
J^0&(\bald{k},\omega) = \frac{\alpha_s ({Q_p^a})^2 m_{\rm D}^2 \gamma \, u^2}{8\pi} \times\\
&\int d^4 x \, e^{i k \cdot x - \rho \, {\rm m} }\left(\frac{1}{(z_-^2 \gamma^2 + \rho^2)^{3/2}} - \frac{\gamma \, u \, z_-}{(z_-^2 \gamma^2 + \rho^2)^2}\right).
\end{split}
\end{equation}
After an integration by parts, the second term in (\ref{fourjvec}) takes the form
\begin{equation}
\int d^4 x \, \frac{u^2 \, e^{i k \cdot x - \rho \, {\rm m} } }{2 (z_-^2 \gamma^2 + \rho^2)}\left( i \bald{k} + {\rm m}(\cos\phi,\sin\phi,0) \right)
\end{equation}
where I am working in plane polar coordinates, $\rho$ and $\phi$, such that $x = \rho \cos \phi$ and $y = \rho \sin \phi$.

It is clear there are three distinct integral forms which need to be evaluated.  After (trivially) integrating out the $t$ dependence to bring down a factor of $2 \pi \delta(\omega - u k_z)$, the three distinct integral forms are
\begin{equation}
\int \frac{d \bald{x} \,e^{- i \bald{k}\cdot \bald{x} -\rho \, {\rm m} }}{(z^2 \gamma^2 + \rho^2)}
\begin{bmatrix}
(1,\cos \phi, \sin\phi)\\
\frac{z}{(z^2 \gamma^2 + \rho^2)}\\
\frac{1}{\sqrt{z^2 \gamma^2 + \rho^2}}
\end{bmatrix}
\equiv
\begin{bmatrix}
\Lambda_1\\
\Lambda_2\\
\Lambda_3
\end{bmatrix}.
\end{equation}
The exponential dependence upon the variable $\phi$ is in the term $-i \rho(k_x \cos{\phi} + k_y \sin{\phi})$.  Re-writing this term as $-i \rho k_T \cos{[\phi - \alpha]}$, where $k_x = k_T \cos \alpha$ and $k_y = k_T \sin \alpha$, the $\phi$ integration can be done using the relations
\begin{equation}\label{bessel}
\begin{split}
\int_0^{2 \pi} \frac{d \phi}{2 \pi} \begin{bmatrix}
1\\
\cos\phi\\
\sin\phi\\
\end{bmatrix} &\exp{[\pm i k_T \rho (\cos{[\phi-\alpha]})]} = \\
&\begin{bmatrix}
J_0(\rho k_T)\\
\pm i J_1(\rho k_T) \cos\alpha\\
\pm i J_1(\rho k_T) \sin\alpha\\
\end{bmatrix}
\end{split}
\end{equation}
where $J_i(x)$ is the Bessel function of the first kind of order $i$.  The final result for $\Lambda_2$ is obtained by using the relation
\begin{equation}
\begin{split}
\label{lamlam2}
\int_{-\infty}^{\infty}d z \int_0^\infty d\rho & \frac{e^{-i z k_z -\rho \, {\rm m} }J_0(\rho k_T) \,z \,\rho}{(\rho^2 + \gamma^2 z^2)^2} = \\
&-\frac{i k_z \pi }{2 \gamma ^2 \sqrt{k_z^2+\left(k_T^2 + {\rm m}^2\right) \gamma^2 + 2 {\rm m} \gamma  |k_z|}}
\end{split}
\end{equation}
which gives, in the large $\gamma$ limit,
\begin{equation}
\label{lamlam2final}
\Lambda_2 = -\frac{(2 \pi) \, i k_z \pi }{2 \gamma^3 \sqrt{k_T^2 + {\rm m}^2}}.
\end{equation}

The first component of $\Lambda_1$, denoted by ${\Lambda_1}_a$, requires evaluating
\begin{equation}
\begin{split}
\label{lamlam1a}
\int_{-\infty}^{\infty}d z \int_0^\infty d\rho & \frac{e^{-i z k_z -\rho \, {\rm m} }J_0(\rho k_T) \,\rho}{(\rho^2 + \gamma^2 z^2)} = \\
&\frac{\pi }{\sqrt{k_z^2+\left(k_T^2 + {\rm m}^2\right) \gamma^2 + 2 {\rm m} \gamma  |k_z|}}.
\end{split}
\end{equation}
The second two components of $\Lambda_1$, denoted by ${\Lambda_1}_b$, can be determined after evaluating
\begin{equation}
\begin{split}
\label{lamlam1b}
\int_{-\infty}^{\infty}d z &\int_0^\infty d\rho  \frac{e^{-i z k_z -\rho \,{\rm m} }J_1(\rho k_T) \,\rho}{(\rho^2 + \gamma^2 z^2)} = \\
&\frac{\pi }{k_T \gamma}\left(1 - \frac{{\rm m}\gamma + |k_z|}{\sqrt{k_z^2+\left(k_T^2 + {\rm m}^2\right) \gamma^2 + 2 {\rm m} \gamma  |k_z|}}\right).
\end{split}
\end{equation}
Again, working in the large $\gamma$ limit, this gives for $\Lambda_1$
\begin{equation}
\label{lamlam1final}
\begin{split}
{\Lambda_1}_a &= \frac{2 \pi^2}{\gamma \sqrt{k_T^2 + {\rm m}^2}} \\
{\Lambda_1}_b &= -\frac{2 \pi^2 i}{\gamma k_T } \left(1 - \frac{{\rm m}}{\sqrt{k_T^2 + {\rm m}^2}}\right)(\cos\alpha,\sin\alpha).
\end{split}
\end{equation}

To determine $\Lambda_3$ it is necessary to evaluate
\begin{equation}\label{trickylam3}
\int_{-\infty}^{\infty}d z \int_0^\infty d\rho \frac{e^{-i z k_z -\rho \, {\rm m} }J_0(\rho k_T) \,\rho}{(z^2 \gamma^2 + \rho^2)^{3/2}}.
\end{equation}
The above form of (\ref{trickylam3}) is difficult to evaluate analytically.  However, it can be made more manageable by including the screening factor in the $z$, rather than $\rho$, integration.  In this case, one has in the large $\gamma$ limit
\begin{equation}
\begin{split}
\int_{-\infty}^{\infty}d z \int_0^\infty d\rho &\frac{e^{-i z k_z -\gamma |z| \, {\rm m} }J_0(\rho k_T) \,\rho}{(z^2 \gamma^2 + \rho^2)^{3/2}} = \\
& \frac{2}{\gamma} \, G_0\left(\frac{{\rm m} + k_T}{2 \sqrt{E_p T}}\right)
\end{split}
\end{equation}
where $z_{min} = (2 \gamma \sqrt{E_p T})^{-1}$ has been used as a short distance cutoff in the $z$ integration.  This gives for $\Lambda_3$
\begin{equation}
\Lambda_3 = \frac{4 \pi}{\gamma} \, G_0\left(\frac{{\rm m} + k_T}{2 \sqrt{E_p T}}\right).
\end{equation}

It is now possible to write down the final result for (\ref{fourjnot}) and (\ref{fourjvec}).  Remembering to include the factor of $2 \pi \delta(\omega - u k_z)$ from the $t$ integration, one has
\begin{equation}
\begin{split}
J^0(\bald{k},\omega) &= \frac{\alpha_s ({Q_p^a})^2 m_{\rm D}^2 u^2}{2} (2 \pi) \delta(\omega - u k_z) \\
&\times\left(\, G_0\left(\frac{{\rm m} + k_T}{2 \sqrt{E_p T}}\right) + \frac{i \pi (\bald{u}\cdot \bald{k})}{4 \gamma \sqrt{k_T^2 + {\rm m}^2}} \right) \\
\bald{J}(\bald{k},\omega) &= \frac{\alpha_s ({Q_p^a})^2 m_{\rm D}^2 u^2}{2} (2 \pi) \delta(\omega - u k_z) \\
&\times \left(\bald{u}\, G_0\left(\frac{{\rm m} + k_T}{2 \sqrt{E_p T}}\right) \right.\\
&\left.+ \frac{i \pi}{4 \gamma}\left(\frac{\sqrt{k_T^2 + {\rm m}^2} - {\rm m}}{k_T^2}\right)\left(u^2\bald{k} +\frac{\bald{u}(\bald{u}\cdot\bald{k}){\rm m}}{\sqrt{k_T^2 + {\rm m}^2}}\right)\right)
\end{split}
\end{equation}
which is the result quoted in (\ref{fourfullsource}).

\end{document}